\documentclass[12pt]{article}
\usepackage{epsfig}
\usepackage{amssymb}
\usepackage{amsmath}
\usepackage{amsfonts}
\usepackage{graphicx}
\usepackage{mathrsfs}
\usepackage[dvips]{color}
\usepackage{multirow}

\newcommand{\vp}{{\vec p}}

\newcommand{\vd}{{\vec \partial}}

\newcommand{\bsigma}{\boldsymbol{\sigma}}

\newcommand{\bnabla}{\boldsymbol{\nabla}}

\newcommand{\bOmega}{\boldsymbol{\Omega}}

\newcommand{\R}{\mathbb{R}}
\newcommand{\C}{\mathbb{C}}

\newcommand{\fz}{\mathfrak{z}}

\newcommand{\fK}{\mathfrak{K}}

\newcommand{\bbe}{\mathbf{e}}

\newcommand{\bbr}{\mathbf{r}}

\newcommand{\bB}{\mathbf{B}}
\newcommand{\bC}{\mathbf{C}}
\newcommand{\bD}{\mathbf{D}}
\newcommand{\bE}{\mathbf{E}}
\newcommand{\bH}{\mathbf{H}}
\newcommand{\bI}{\mathbf{I}}
\newcommand{\bJ}{\mathbf{J}}

\newcommand{\bM}{\mathbf{M}}

\newcommand{\bS}{\mathbf{S}}

\newcommand{\cK}{\mathcal{K}}

\newcommand{\cP}{\mathcal{P}}

\newcommand{\be}{\begin{equation}}
\newcommand{\ee}{\end{equation}}
\newcommand{\bea}{\begin{eqnarray}}
\newcommand{\eea}{\end{eqnarray}}
\newcommand{\nn}{\nonumber}
\newcommand{\kt}{\rangle}
\newcommand{\br}{\langle}

\newcommand{\ed}{\end{document}}

\newcommand{\bi}{\begin{itemize}}
\newcommand{\ei}{\end{itemize}}

\newcommand{\emp}{\varnothing}
\newcommand{\bce}{\begin{center}}
\newcommand{\ece}{\end{center}}

\newcommand{\sD}{\mathscr{D}}

\newcommand{\sT}{\mathscr{T}}

\newcommand{\RE}{{\rm Re}}
\newcommand{\IM}{{\rm Im}}

\begin{document}

\title{Unidirectional Invisibility and Nonreciprocal Transmission\\ in Two and Three Dimensions}

\author{Farhang Loran\thanks{E-mail address:loran@cc.iut.ac.ir}  
~and Ali~Mostafazadeh\thanks{E-mail address:
amostafazadeh@ku.edu.tr}\\[6pt]
$^*$Department of Physics, Isfahan University of Technology, \\ Isfahan 84156-83111, Iran\\[6pt]
$^\dagger$Departments of Mathematics and Physics, Ko\c{c} University,\\  34450 Sar{\i}yer,
Istanbul, Turkey}

\date{ }
\maketitle

\begin{abstract}

We explore the phenomenon of unidirectional invisibility in two dimensions, examine its optical realizations, and discuss its three-dimensional generalization. In particular we construct an infinite class of unidirectionally invisible optical potentials that describe the scattering of normally incident transverse electric waves by an infinite planar slab with refractive-index modulations along both the normal directions to the electric field. A by-product of this investigation is a demonstration of nonreciprocal transmission in two dimensions. To elucidate this phenomenon we state and prove a general reciprocity theorem that applies to quantum scattering theory of real and complex potentials in two and three dimensions.
\vspace{2mm}

\vspace{2mm}


\end{abstract}

\section{Introduction}

In one dimension a real scattering potential has identical reflection and transmission coefficients for incoming waves from the left and right, i.e., it obeys reciprocity in reflection and transmission. In contrast, a complex scattering potential can violate reciprocity in reflection \cite{jpa-2014c}. A remarkable consequence of this observation is the existence of unidirectionally reflectionless and invisible potentials \cite{invisible1,lin,invisible2,invisible3,pra-2013a,pra-2014a,pra-2015c,invisible4,invisible5}. These have recently attracted a great deal of attention, mainly because they model certain one-way optical devices with possible applications in optical circuitry.

One of the most sought-after examples of a one-way element of an optical circuit is an optical diode \cite{jalas}. It is generally believed that real and complex potentials are incapable of modeling an optical diode, because they are bound to respect reciprocity in transmission \cite{ahmed,prl-2009,sanchez}. The same problem arises also in acoustics where designing acoustic waveguides with nonreciprocal transmission is a major area of research \cite{Fleury-2015}.

The search for means of achieving nonreciprocal transmission is dominated by the use of nonlinear, time-dependent, or magnetic materials \cite{aplet,jalas,Fleury-2015}. The present work is motivated by the observation that reciprocity in reflection and transmission can both be violated within the confines of linear, stationary, isotropic, nonmagnetic material provided that one employs a genuine multidimensional setup. In particular, it is possible to construct unidirectionally invisible potentials in two and three dimensions that enjoy nonreciprocal transmission.

The basic conceptual framework for the present work is the multidimensional transfer matrix formulation of scattering theory that we report in \cite{p129}. We use it to offer a precise and convenient way of describing unidirectional reflection, unidirectional invisibility, and transmission reciprocity in dimensions higher than one.

The organization of this article is as follows. In Sec.~\ref{Sec2} we review the formulation of scattering theory presented in \cite{p129} and use it to give a precise definition for the notions of unidirectional reflectionlessness, unidirectional transparency, unidirectional invisibility, and reciprocal transmission in two dimensions. Here we also derive some useful relations characterizing these notions. In Sec.~\ref{Sec3} we prove a general reciprocity theorem that applies to real and complex scattering potentials in two dimensions. In Sec.~\ref{Sec4}, we construct an infinite class of unidirectionally invisible potentials in two dimensions that possess nonreciprocal transmission. In Sec.~\ref{Sec5} we examine an optical realization of a particular example of these potentials and discuss the physical implications of their unidirectional invisibility and nonreciprocal transmission properties. In Sec.~\ref{Sec6} we generalize the results of the preceding sections to three dimensions, in particular establishing the three-dimensional analog of the reciprocity theorem of Sec.~\ref{Sec3} and constructing finite-range unidirectionally invisible potentials with nonreciprocal transmission in three dimensions. Finally, in Sec.~\ref{Sec7} we summarize our findings and present our concluding remarks.

\section{Transfer Matrix and Unidirectional Invisibility in 2D}
\label{Sec2}

Consider the Schr\"odinger equation
    \be
    -\boldsymbol{\nabla}^2\psi(\bbr)+v(\bbr)\psi(\bbr)=k^2\psi(\bbr),
    \label{sch-eq}
    \ee
for a real or complex scattering potential $v$, where $\bbr:=(x,y)$ and $k$ is the wave number. Let us identify the $x$-axis with the scattering axis, and recall that for $x\to\pm\infty$ the scattering solutions of (\ref{sch-eq}) have the form
    \be
    \frac{1}{2\pi}
	\int_{-k}^k dp\, e^{ipy}\left[A_\pm(p)e^{i\omega(p)x}+
	B_\pm(p) e^{-i\omega(p)x}\right],
    \label{asym}
	\ee
where $A_\pm(p)$ and $B_\pm(p)$ are coefficient functions vanishing for $p\notin[-k,k]$ and $\omega(p):=\sqrt{k^2-p^2}$, \cite{p129}. By definition, the transfer matrix of $v$ is the $2\times 2$ matrix operator $\bM(p)$ satisfying
    \be \left[\begin{array}{c}
	A_+(p) \\ B_+(p) \end{array}\right]=\bM(p)\left[\begin{array}{c}
	A_-(p) \\ B_-(p) \end{array}\right].
	\label{M-def}
	\ee
Similarly to its one-dimensional analog \cite{ap-2014}, it can be expressed as a time-ordered exponential \cite{p129};
    \begin{align}
    &\bM(p):=\sT\exp\int_{-\infty}^\infty -i\bH(x,p)dx,
    \label{M=}
    \end{align}
where $x$ plays the role of time, $\bH(x,p)$ is the non-Hermitian effective Hamiltonian operator:
    \begin{align}
    &\bH(x,p):=\frac{1}{2\omega(p)} e^{-i\omega(p)x\boldsymbol{\sigma}_3}
	v(x,i\partial_p)\,\boldsymbol{\cK}\,e^{i\omega(p)x\boldsymbol{\sigma}_3},
    \label{H=}
    \end{align}
$\boldsymbol{\sigma}_i$ are the Pauli matrices, $\boldsymbol{\cK}:=\boldsymbol{\sigma}_3+i\boldsymbol{\sigma}_2$, $v(x,i\partial_p)$ is the integral operator acting on test functions $\phi:[-k,k]\to\C$ according to
    \[v(x,i\partial_p)\phi(p):=\frac{1}{2\pi}\int_{-k}^k dq\,\tilde v(x,p-q) \phi(q),\]
and $\tilde v(x,\fK_y)$ denotes the Fourier transform of $v(x,y)$ with respect to $y$, i.e., $\tilde v(x,\fK_y):=\int_{-\infty}^\infty dy\,e^{-i\fK_y y}v(x,y)$.

The transfer matrix $\bM(p)$ contains all the scattering information about $v$. To see this we recall that for a left-incident plane wave with a momentum $\mathbf{k}$ pointing along the positive $x$-axis,
the scattering solutions $\psi^{\rm l}$ of  (\ref{sch-eq}) have the asymptotic form:
    \be
    \psi^{\rm l}(\bbr)=e^{ikx}+\sqrt{\frac{i}{kr}}\,e^{ikr} f^{\rm l}(\theta)~~~~{\rm as}~r\to\infty,
    \label{psi-left}
    \ee
where $(r,\theta)$ are the polar coordinates of $\bbr$, and $f^{\rm l}(\theta)$ is the scattering amplitude for the left-incident waves. In order to express the latter in terms of $\bM(p)$, we introduce
    \begin{align}
    &T^{\rm l}_-(p):=B_-(p), &&T^{\rm l}_+(p):=A_+(p)-A_-(p),
    \label{T-def-L}
    \end{align}
and note that for a left-incident wave,
    \begin{align}
	&A_-(p)=2\pi\delta(p), && B_+(p)=0,
	\label{scat-L}
	\end{align}
where $\delta$ stands for the Dirac delta-function. In Ref.~\cite{p129}, we show that (\ref{M-def}), (\ref{T-def-L}), and (\ref{scat-L}) imply
    \begin{align}
	&T^{\rm l}_-(p)=-2\pi M_{22}^{-1}M_{21}\delta(p),
    \label{Tm-L}\\
    &T^{\rm l}_+(p)=2\pi(M_{11}-1-M_{12}M_{22}^{-1}M_{21})\delta(p),
    \label{Tp-L}\\
	&f^{\rm l}(\theta)=-(2\pi)^{-\frac{1}{2}} i k|\cos\theta|\,T^{\rm l}_\pm(k\sin\theta),
 	\label{f-L}
    \end{align}
where $M_{ij}$ are the entries of $\bM(p)$ and the $\pm$ in (\ref{f-L}) stands for ${\rm sgn}(\cos\theta)$.

Similarly, we can introduce the scattering amplitude $f^{\rm r}(\theta)$ for a right-incident wave by expressing the asymptotic form of the corresponding scattering solution as
    \be
    \psi^{\rm r}(\bbr)=e^{-ikx}+\sqrt{\frac{i}{kr}}\,e^{ikr} f^{\rm r}(\theta)~~~~{\rm as}~r\to\infty.
    \label{psi-right}
    \ee
To relate $f^{\rm r}(\theta)$ to $\bM(p)$, we define
    \begin{align}
    &T^{\rm r}_-(p):=B_-(p)-B_+(p), &&T^{\rm r}_+(p):=A_+(p),
    \label{T-def-R}
    \end{align}
and set
    \begin{align}
	&A_-(p)=0 , && B_+(p)=2\pi\delta(p).
	\label{scat-R}
	\end{align}
These together with (\ref{M-def}) give
    \begin{align}
	&T^{\rm r}_-(p)=-2\pi (1-M_{22}^{-1})\delta(p),
    \label{Tm-R}\\
    &T^{\rm r}_+(p)=2\pi M_{12}M_{22}^{-1}\delta(p),
    \label{Tp-R}\\
	&f^{\rm r}(\theta)=-(2\pi)^{-\frac{1}{2}} i k|\cos\theta|\,T^{\rm r}_\pm(k\sin\theta),
    \label{f-R}
    \end{align}
where again $\pm:={\rm sgn}(\cos\theta)$.

It is important to note that $M_{ij}$ and $T^{\rm l/r}_{\pm}(p)$ depend on the wavenumber $k$. Therefore any quantity or concept whose definition involves $M_{ij}$ or $T^{\rm l/r}_{\pm}(p)$ is $k$-dependent. 
    \begin{itemize}
    \item[]{\em Definition~1.} Let $v:\R^2\to\C$ be a scattering potential and $k$ be a wavenumber. Choose a Cartesian coordinate system in which the scattering axis coincides with the $x$-axis. Then $v$ is said to be 
    \begin{itemize}
    \item[-] left- (respectively right-) reflectionless if $f^{\rm l}(\theta)=0$ for all $\theta\in(\frac{\pi}{2},\frac{3\pi}{2})$ (respectively $f^{\rm r}(\theta)=0$ for all $\theta\in (-\frac{\pi}{2},\frac{\pi}{2})$), 
    \item[-] left- (respectively right-) transparent if $f^{\rm l}(\theta)=0$ for all $\theta\in (-\frac{\pi}{2},\frac{\pi}{2})$ (respectively $f^{\rm r}(\theta)=0$ for all $\theta\in (\frac{\pi}{2},\frac{3\pi}{2})$), 
    \item[-] left- (respectively right-) invisible if it is both left- (respectively right-) reflectionless and transparent. 
    \item[-]  unidirectionally reflectionless (respectively transparent or invisible) if it is either left- or  right-reflectionless (respectively transparent or invisible), but not both.
    \end{itemize}
    \end{itemize}
The following is a direct consequence of (\ref{f-L}) and (\ref{f-R}). 
    \begin{itemize}
    \item[]{\em Theorem~1.} Let $v$ be as in Definition~1. Then $v$ is
    \begin{itemize}
    \item[-] left- (respectively right-) reflectionless if and only if $T^{\rm l}_-(p)=0$ (respectively $T^{\rm r}_+(p)=0$) for all $p\in[-k,k]$, 
    \item[-] left- (respectively right-) transparent if $T^{\rm l}_+(p)=0$ (respectively $T^{\rm r}_-(p)=0$) for all $p\in[-k,k]$.
    \end{itemize}
    \end{itemize}
In view of (\ref{Tm-L}), (\ref{Tp-L}), (\ref{Tm-R}), and (\ref{Tp-R}), we also have the following useful result.
    \begin{itemize}
    \item[]{\em Theorem~2.} Let $v$ be as in Definition~1. Then $v$ is 
    \begin{itemize}
    \item[-] left-reflectionless, if and only if $M_{21}\delta(p)=0$ for all $p\in[-k,k]$.
    \item[-] left-invisible, if and only if $M_{21}\delta(p)=(M_{11}-1)\delta(p)=0$ for all $p\in[-k,k]$.
    \item[-] right-invisible, if and only if $M_{12}\delta(p)=(M_{22}-1)\delta(p)=0$ for all $p\in[-k,k]$.
    \end{itemize}
    \end{itemize}

\section{Reciprocity Principle in 2D}
\label{Sec3}

The phrase `Reciprocity Principle' is usually taken to be synonymous to the `Lorentz  Reciprocity Theorem' for electromagnetic fields which follows from Maxwell's equations \cite{jalas}. Here we formulate a reciprocity principle that applies directly to the quantum scattering theory as defined by the Schr\"odinger equation~(\ref{sch-eq}). First, we define what we mean by reciprocity in transmission in two dimensions. 
    \begin{itemize}
    \item[]{\em Definition~2.} A real or complex scattering potential $v$ is said to have reciprocal transmission for a wavenumber $k$, if for all $p\in[-k,k]$, $T_+^{\rm l}(p)=T_-^{\rm r}(p)$.  According to (\ref{Tp-L}) and (\ref{Tm-R}), this is equivalent to
            \be
            (M_{11}-M_{22}^{-1}-M_{12}M_{22}^{-1}M_{21})\delta(p)=0.
            \label{non-rec-trans}
            \ee
        We can also use (\ref{f-L}) and (\ref{f-R}) to state it in the form
	       \be
	       f^{\rm l}(\theta)=f^{\rm r}(\pi-\theta)~~{\rm for~all}~~\theta\in
            \mbox{$(-\frac{\pi}{2},\frac{\pi}{2})$}.
	       \label{f-rec}
	       \ee
    \end{itemize}

The content of Definitions~1 and~2 reduce to their well-known one-dimensional analog provided that we replace $\delta(p)$ with 1 and identify $M_{ij}$ with the entries of the transfer matrix in one dimension \cite{pra-2013a}. In this case, Condition~(\ref{non-rec-trans}) for reciprocal transmission reduces to $M_{11}-M_{22}^{-1}-M_{12}M_{22}^{-1}M_{21}=0$ which is equivalent to $\det\bM=1$. But in one-dimension, one can use Abel's theorem for linear homogeneous ordinary differential equations \cite{abel} to show that `$\det\bM=1$' holds for every real or complex scattering potential, \cite{jpa-2009}. This proves that nonreciprocal transmission is forbidden in one dimension. It also suggests that in order to establish (\ref{non-rec-trans}) or its violation in two (and higher) dimensions, we should seek for an analog of Abel's theorem for the Schr\"odinger equation (\ref{sch-eq}), which is a partial differential equation.

In one dimension, the Schr\"odinger equation reads: $-\psi''(x)+v(x)\psi(x)=k^2\psi(x)$, and Abel's theorem states that the Wronskian, $\psi_1(x)\psi_2'(x)-\psi_2(x)\psi_1'(x)$, of any pair of solutions of this equation, $\psi_1$ and $\psi_2$, is $x$-independent. A higher-dimensional generalization of the Wronskian is the vector field,
	$\bJ(\bbr):=\psi_1(\bbr)\bnabla\psi_2(\bbr)-\psi_2(\bbr)\bnabla\psi_1(\bbr)$.
It is easy to check that whenever $\psi_1$ and $\psi_2$ are solutions of the Schr\"odinger equation~(\ref{sch-eq}), $\bJ$ is divergence-free;
	\be
	\bnabla\cdot\bJ(\bbr)=0.
	\label{div-J}
	\ee
	
Now, let $a$ and $b$ be positive real numbers, $V$ be the rectangular region enclosed between the planes $x=0$, $x=a$, and $y=\pm b$, and
	\be
	j(x):=\int_{-\infty}^\infty\!\! dy\,\left[\psi_1(\bbr)\partial_x\psi_2(\bbr)-\psi_2(\bbr)\partial_x\psi_1(\bbr)\right].
	\label{j=}
	\ee
Suppose that $\partial_y\psi_j(x,y)\to 0$ for $y\to\pm\infty$, which is consistent with (\ref{psi-left}) and (\ref{psi-right}). Then applying the divergence theorem to $\bJ(\bbr)$ and $V$, using (\ref{div-J}), and taking the $b\to\infty$ limit, we find that $j(a)=j(0)$. Because $a$ is arbitrary, this means that $j(x)$ does not depend on $x$. In particular, we have\footnote{Here we assume that $j(x)$ has finite asymptotic values for $x\to\pm\infty$. For our purposes we only need to consider $j(x)$ for the the scattering solutions $\psi^{\rm l/r}$, and for these $\lim_{x\to\pm\infty}j(x)$ exist.}
	 \be
	j(-\infty)=j(\infty).
	\label{j=j}
	\ee
	
Because scattering solutions of (\ref{sch-eq}) admit an asymptotic expression of the form (\ref{asym}), we can calculate $j(\pm\infty)$ whenever $\psi_1$ and $\psi_2$ are scattering solutions. Let $A_{j\pm}$ and $B_{j\pm}$ respectively denote the coefficient functions $A_\pm$ and $B_\pm$ appearing in the asymptotic expression (\ref{asym}) for $\psi_j$. Then substituting this expression in
(\ref{j=}), we obtain
	\begin{align}
	&j(x)=\frac{-i}{\pi}\int_{-k}^kdp\, \omega(p)\Delta_\pm(p)~~~{\rm for}~~x\to\pm\infty,
	\label{j=2}\\
	&\Delta_\pm(p):=A_{1\pm}(-p)B_{2\pm}(p)-B_{1\pm}(-p)A_{2\pm}(p).
	\label{D=}
	\end{align}
The fact that the right-hand side of (\ref{j=2}) does not involve $x$ provides an independent verification of (\ref{j=j}).
	
Next we use the definition of the transfer matrix, namely (\ref{M-def}), to express the $A_{j+}$ and $B_{j+}$ appearing in (\ref{D=}) in terms of $A_{j-}$ and $B_{j-}$. Inserting the resulting formula in (\ref{j=2}), we can express (\ref{j=j}) in the form
	\bea
	&&\int_{-k}^k\!\!dp\,\omega(p)
	\left[\bC_{1+}(-p)^T\bOmega\bC_{2+}(p)\right.\nn\\
	&&\quad\quad\quad\quad\quad~\left. -\bC_{1-}(-p)^T\bOmega\bC_{2-}(p)\right]=0,
	\label{CC}
	\eea
where the superscript `$T$' stands for the transpose,
	\begin{align}
	&\bC_{j-}(p):=\left[\begin{array}{c}
	A_{j-}(p)\\ B_{j-}(p)\end{array}\right],
	&& \bC_{j+}(p):=\bM(p)\bC_{j-}(p),
	\label{C=}
	\end{align}
and $\bOmega:=i\bsigma_2=\left[\begin{array}{cc}
0 & 1\\ -1 & 0\end{array}\right]$ is the standard $2\times 2$ symplectic matrix.
Because the choice of $\bC_{j-}(p)$ is arbitrary, (\ref{CC}) is equivalent to
	\be
	\overleftarrow{\bM(-p)^T}\bOmega\,\overrightarrow{\bM(p)}=\bOmega.
	\label{rec1}
	\ee
Here we use arrows to stress that $\bM(-p)^T$ acts on the test functions appearing to its left while $\bM(p)$ acts on those appearing to its right.

In order to elucidate the physical meaning of Eq.~(\ref{j=j}), which is equivalent to (\ref{CC}) and (\ref{rec1}), we identify $\psi_1$ and $\psi_2$ with scattering solutions $\psi^{\rm l}$ and $\psi^{\rm r}$ associated with the incident plane waves from the left and right, respectively. This corresponds to setting
	\begin{align*}
	&A_{1-}(p)=2\pi\delta(p), &&B_{1+}(p)=0,\\
	&B_{2+}(p)=2\pi\delta(p),&&A_{2-}(p)=0.
	\end{align*}
Substituting these relations in (\ref{D=}) and making use of (\ref{scat-L}), (\ref{scat-R}), and (\ref{j=2}), we can write (\ref{j=j}) as
	\be
	T^{\rm l}_+(0)=T^{\rm r}_-(0),
	\label{rec2}
	\ee
or $f^{\rm l}(0)=f^{\rm r}(\pi)$. In view of (\ref{f-L}) and (\ref{f-R}) this proves the following reciprocity theorem.
    \begin{itemize}
    \item[]{\em Theorem~3 (Reciprocity Principle).} For every real or complex scattering potential, the forward scattering amplitude for the left- and right-incident waves coincide.
    \end{itemize}

\noindent Notice that this theorem does not prohibit nonreciprocal transmission in the sense of Definition~2. In particular, as we show in the next section there are scattering potentials with different total transmission cross section from the left and right.

\section{Unidirectionally Invisible Potentials with Nonreciprocal Transmission}
\label{Sec4}

The recent interest in unidirectional invisibility is initiated by the study of the locally periodic optical potentials of the form  \cite{invisible1,lin,invisible2}
	\be
	v(x)=\fz\, e^{iKx} \chi_{a}(x) =
	\left\{\begin{array}{cc}
	\fz\, e^{iKx} & {\rm for}~x\in[0,a],\\[3pt]
	0 &{\rm otherwise},\end{array}\right.
	\label{exp}
	\ee
where $\fz$ is a coupling constant, $K:=2\pi/a$, $a$ is a positive real parameter, and
$\chi_{a}$ is the characteristic function for the interval $[0,a]$, i.e.,
	\[\chi_{a}(x):=\left\{\begin{array}{cc}
	1 & {\rm for}~x\in[0,a],\\[3pt]
	0 & {\rm otherwise}.\end{array}\right.\]
This potential is unidirectionally invisible for $k=\pi/a$ provided that $|\fz|$ is so small that the scattering properties of $v$ can be determined using the first Born approximation, i.e., it displays perturbative unidirectional invisibility \cite{pra-2014a}. In optical applications, where $v$ is related to the relative permittivity ${\hat\varepsilon}$ of the material according to $v(x)=k^2[1-{\hat\varepsilon}(x)]$, this condition holds whenever $|{\hat\varepsilon}(x)-1|\ll 1$.

In one dimension one can use the dynamical formulation of scattering theory developed in Ref.~\cite{ap-2014} to construct scattering potentials that enjoy exact unidirectional invisibility \cite{ap-2014,pra-2014b}. These have a more complicated structure than (\ref{exp}). Reference~\cite{pra-2014a} provides a complete characterization of finite-range potentials that similarly to (\ref{exp}) support perturbative unidirectional invisibility. In the remainder of this section, we construct a class of perturbative unidirectionally invisible potentials in two dimensions that generalize (\ref{exp}).

Consider a planar slab of optical material with relative permittivity ${\hat\varepsilon}$
that is located between the planes $x=0$ and $x=a$  in three-dimensional Euclidean space.  Suppose that the slab has translational symmetry along the $z$-directions, so that ${\hat\varepsilon}$ depends only on $x$ and $y$. Then the interaction of this slab with the time-harmonic normally incident $z$-polarized transverse electric waves of the form
	\be
    \bE(x,y,z)=E_0\,e^{-ik{\rm c}t}\psi(x,y)\bbe_z
    \label{E=ez}
    \ee
is described by the Schr\"odinger equation~(\ref{sch-eq}) with the potential:
	\be
	v(x,y)=k^2[1-{\hat\varepsilon}(x,y)] \chi_{a}(x).
	\label{exp-v}
	\ee
In Eq.~(\ref{E=ez}), $E_0$ is a nonzero complex parameter, ${\rm c}$ is the speed of light in vacuum, and $\bbe_z$ is the unit vector pointing along the positive $z$-axis.

In what follows we consider situations where $|{\hat\varepsilon}(x,y)-1|$ is so small that the first Born approximation provides a reliable description of the behavior of the system. In this case, we can compute the transfer matrix (\ref{M=}) using the formula
	\be
	\bM(p)\approx\bI-i\int_{-\infty}^\infty dx\,\bH(x,p),
	\label{BA}
	\ee
where $\bI$ is the $2\times 2$ identity matrix, and `$\approx$' stands for the first Born approximation. Substituting (\ref{H=}) in this relation to determine $M_{ij}$ and using these in (\ref{Tm-L}), (\ref{Tp-L}), (\ref{Tm-R}), and (\ref{Tp-R}) give
	\begin{align}
	&T^{\rm l}_\pm(p)\approx\frac{-i}{2\omega(p)}\,\tilde{\tilde v}(-p_\mp,p),
	\label{Tpm-L=}\\
	&T^{\rm r}_\pm(p)\approx \frac{-i}{2\omega(p)}\,\tilde{\tilde v}(p_\pm,p),
	\label{Tpm-R=}
	\end{align}
where $\tilde{\tilde v}(\fK_x,\fK_y)$ is the two-dimensional Fourier transform of $v(x,y)$, i.e.,
	\be
	\tilde{\tilde v}(\fK_x,\fK_y):=
	\int_{-\infty}^\infty \!\!dx\int_{-\infty}^\infty\!\! dy\,e^{-i(\fK_xx+\fK_yy)}v(x,y),
	\label{FT-v}
	\ee
and $p_\pm:=k\pm\omega(p)=k\pm\sqrt{k^2-p^2}$.

It is easy to see that Eqs.~(\ref{Tpm-L=}) and (\ref{Tpm-R=}) are consistent with the statement of the Reciprocity Principle (Theorem~3); clearly for $p=0$, we have $p_-=0$, and (\ref{rec2}) holds.

In view of Definition~1, Eqs.~(\ref{Tpm-L=}) and (\ref{Tpm-R=}) imply that $v$ is unidirectionally invisible from the right if and only if the following conditions hold.
	\begin{align}
	&\tilde{\tilde v}(p_\pm,p)=0~~{\rm for~all}~~p\in[-k,k],
	\label{R-inv}\\
	&|\tilde{\tilde v}(-p_-,p)|+|\tilde{\tilde v}(-p_+,p)|\neq 0~~~{\rm for~some}~~p\in[-k,k].~~
	\nn 
	\end{align}
In order to obtain explicit examples of potentials of the form (\ref{exp-v}) that satisfy these conditions, we expand them in their Fourier series in $[0,a]$. This gives
	\be
	v(x,y)=\chi_{a}(x)\sum_{n=-\infty}^\infty c_n(y)e^{inKx},
	\label{FS}
	\ee
where $c_n(y):=\frac{1}{a}\int_0^a dx\, e^{-inKx}v(x,y)$ and
	\begin{align}
	&K:=\frac{2\pi}{a}.
	\label{cn=}
	\end{align}
	
In view of  (\ref{FT-v})  and (\ref{FS}),
	\be
	\tilde{\tilde v}(\fK_x,\fK_y)=\sum_{n=-\infty}^\infty
	\left[\frac{e^{ia(nK-\fK_x)}-1}{i(nK-\fK_x)}\right]\tilde c_n(\fK_y).
	\label{ttv=}
	\ee
Substituting this relation in (\ref{R-inv}), we find two complex equations for the coefficient functions $\tilde c_n(p)$. These characterize perturbative right-invisible potentials $v(x,y)$ that vanish for $x\notin[0,a]$. To obtain concrete examples of such potentials, we take a pair of distinct nonzero integers, $\ell$ and $m$, and set
	\be
	c_n(y)=0~~~\mbox{for}~~~ n\neq 0,\ell,m.
	\label{cn=0}
	\ee
We then solve (\ref{R-inv}) to express  $\tilde c_0$, $\tilde c_\ell$, and $\tilde c_m$ in terms of an arbitrary function that we denote by $\tilde g$. This gives
	\bea
	\tilde c_0(p)&=&p^2 \tilde g(p),
	\label{tc-0}\\
 	\tilde c_\ell(p)&=&
	\frac{m\left[\ell K(\ell K-2k)+p^2\right]\tilde g(p)}{\ell-m},
	\label{tc-L}\\
	\tilde c_m(p)&=&
	\frac{\ell \left[m K(m K-2k)+p^2\right]\tilde g(p)}{m-\ell}.
 	\label{tc-m}
 	\eea	
Evaluating the inverse Fourier transform of these relations and using the result together with
(\ref{cn=0}) in (\ref{FS}), we find
	\bea
	v(x,y)&=&\chi_{a}(x)\Big\{
    -\Big(1+\frac{\ell e^{i\ell K x}-m e^{im K x}}{\ell-m}\Big)g''(y)
	\label{v=g}\\
	&& +\frac{\ell m K}{\ell-m}
	\Big[(\ell K-2k)e^{i\ell K x}-(m K-2k)e^{i m K x}\Big]g(y)\Big\},~~
	\nn
	\eea
where $g$ is the inverse Fourier transform of $\tilde g$.
	
\begin{figure}
	\begin{center}
	\includegraphics[scale=.5]{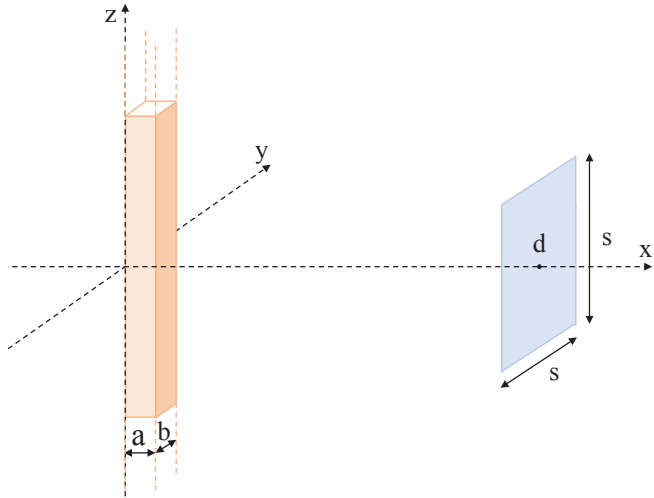}
	\caption{Schematic view of an optically active wire of rectangular cross section with a screen placed to its right at $x=d$.}
	\label{fig1}
	\end{center}
	\end{figure}

Equation~(\ref{v=g}) gives a right-invisible potential for each choice of $k,\ell,m,$ and $g$, provided that $g$ and $g''$ be such that (\ref{v=g}) defines a well-behaved scattering potential and that the first Born approximation is justified. Typical examples are
	\begin{align}
	&g(y)=g_0\, e^{-y^2/2b^2},
	\label{g=exp}\\
	&g(y)=g_0\,b^{-4}y^2(y-b)^2\chi_b(y),
	\label{g=2}
	\end{align}
where $g_0$ is a possibly complex nonzero coupling constant, $b$ is a positive real parameter, and $|g_0|$ is sufficiently small. Notice that (\ref{g=2}) corresponds to a finite-range right-invisible potential whose support is the rectangular region: $[0,a]\times[0,b]$. It describes an optically active wire of rectangular cross section that is aligned along the $z$-axis, as shown in Fig.~\ref{fig1}.

In order to make sure that the potentials (\ref{v=g}) are not invisible from the left, we examine $T^l_\pm(p)$. First, we substitute (\ref{tc-0}) -- (\ref{tc-m}) in (\ref{ttv=}) and make use of (\ref{cn=}) and (\ref{cn=0}) to derive
	\be
	\tilde{\tilde v}(\fK_x,\fK_y)=\frac{\ell m K^2 (1-e^{-ia\fK_x})[\fK_y^2+(\fK_x-2k)\fK_x]\tilde g(\fK_y)}{
	i \fK_x(\fK_x-\ell K)(\fK_x-m K)}.\nn
	\ee
This together with (\ref{Tpm-L=}) imply
	\be
	T^{\rm l}_\pm(p)\approx\left[\frac{2\ell m K^2 k (1-e^{i a p_\mp})}{\omega(p)
	(p_\mp+\ell K)(p_\mp+m K)}\right]\tilde g(p).
	\label{TL=g}
	\ee
This equation provides an explicit demonstration of nonreciprocity in  transmission, because for generic choices of $g$ and $p$, we have $T^l_+(p)\neq 0$, while $T^r_-(p)\approx 0$ for all $p\in[-k,k]$.

Using (\ref{TL=g}) in (\ref{f-L}), we also find
	\be
	f^{\rm l}(\theta)\approx
	\frac{-\sqrt 2\, i \ell m K^2 k \left[1-e^{i ak(1-\cos\theta)}\right]\tilde g(k\sin\theta)}{\sqrt \pi
	[k(1-\cos\theta)+\ell K][k(1-\cos\theta)+m K]}.
	\ee
According to this relation and the fact that $f^{\rm r}(\theta)\approx 0$, the potential (\ref{v=g}) is unidirectionally invisible from the right, but it is neither left-reflectionless nor left-transparent.

\section{Optical Manifestation of Unidirectional Invisibility and Nonreciprocal Transmission}
\label{Sec5}

We can identify the unidirectionally invisible potentials of the form (\ref{v=g}) with optical potentials describing the interaction of the linearly polarized electromagnetic waves (\ref{E=ez}) with an optically active wire of rectangular cross section as depicted in Fig.~\ref{fig1}. In this section we explore the implications of the nonreciprocal transmission
property of these potentials for the transmitted power associated with the left- and right-incident waves. This requires the computation of the Poynting vector for the corresponding scattered waves.

First, we recall that for a charge-free, nonmagnetic, isotropic medium, the time-averaged energy density and Poynting vector are respectively given by
    {\small\begin{align}
    &\br u\kt=\frac{1}{4}\RE(\bE\cdot\bD^*+\bB\cdot\bH^*),
    &&\br\bS\kt:=\frac{1}{2}\RE(\bE\times\bH^*),
    \label{u-S}
    \end{align}}%
where $\bE$ and $\bH$ are the electric and magnetic fields, $\bD=\varepsilon_0\hat{\varepsilon}\,\bE$, $\bB=\mu_0\bH$, and $\varepsilon_0$ and $\mu_0$ are the permittivity and permeability of the vacuum, respectively. For time-harmonic waves,
Maxwell's equations imply that $\bH=\frac{1}{ik}\sqrt{\frac{\varepsilon_0}{\mu_0}}\boldsymbol{\nabla}\times\bE$. Substituting this relation in (\ref{u-S}) and making use of (\ref{E=ez}), we have
    \bea
    \br u\kt&=&\frac{\varepsilon_0|E_0|^2}{4}\left\{\RE[\hat\varepsilon(\bbr)]|\psi(\bbr)|^2+
    k^{-2}|\boldsymbol{\nabla}\psi(\bbr)|^2\right\},~~~~
    \label{u-psi}\\
    \br\bS\kt&=&\frac{|E_0|^2}{2\mu_0 {\rm c}k} \,\IM\left[\psi(\bbr)^*\boldsymbol{\nabla}\psi(\bbr)\right],
    \label{s-psi}
    \eea
where `$\RE$' and `$\IM$' respectively denote the real and imaginary part of their argument.

Now, consider the time-averaged energy density and Poynting vector for the scattered waves $\psi^{\rm l/r}$, which we denote by $\br u^{\rm l/r}\kt$ and $\br\bS^{\rm l/r}\kt$, respectively. We can use the latter to compute the reflected and transmitted power to the left ($x=-\infty)$ and right ($x=\infty)$. It is not difficult to see that the contribution of the scattering potential to these quantities, i.e., the difference between their value in the presence and absence of the potential, is proportional to
    \be
    \Delta\cP_\pm^{\rm l/r}:=\pm \int_{-\infty}^\infty dy\: \bbe_x\cdot(\br\bS^{\rm l/r}\kt-\br\bS_\emp\kt)\Big|_{x=\pm\infty},
    \label{cP-def}
    \ee
where $\bbe_x$ is the unit vector pointing along the positive $x$-axis, and the subscript `$\emp$' means that the corresponding quantity is computed in the absence of the potential, i.e., for $v=0$. Notice that $\Delta\cP_-^{\rm l}$, $\Delta\cP_+^{\rm l}$, $\Delta\cP_-^{\rm r}$, and $\Delta\cP_+^{\rm r}$ respectively correspond to the reflected power for $\psi^l$, transmitted power for $\psi^l$, transmitted power for $\psi^r$, and reflected power for $\psi^r$.

We can use (\ref{asym}), (\ref{T-def-L}), (\ref{scat-L}), (\ref{f-L}), (\ref{T-def-R}), (\ref{scat-R}), (\ref{f-R}), (\ref{s-psi}), and (\ref{cP-def}) to compute $\Delta\cP_\pm^{\rm l/r}$. The result is
    \begin{align}
    \Delta\cP_-^{\rm l}
    &=\frac{|E_0|^2}{2\mu_0ck}\int_{\frac{\pi}{2}}^{\frac{3\pi}{2}} d\theta|f^{\rm l}(\theta)|^2,\nn
    \\
    \Delta\cP_+^{\rm l}
    &=\frac{|E_0|^2}{2\mu_0ck}\left\{\int_{\frac{-\pi}{2}}^{\frac{\pi}{2}} d\theta|f^{\rm l}(\theta)|^2-\sqrt{8\pi}\,\IM[f^{\rm l}(0)]\right\},\nn\\
    \Delta\cP_-^{\rm r}
    &=\frac{|E_0|^2}{2\mu_0ck}\left\{
    \int_{\frac{\pi}{2}}^{\frac{3\pi}{2}} d\theta|f^{\rm r}(\theta)|^2
    -\sqrt{8\pi}\,\IM[f^{\rm r}(0)]\right\},\nn\\
    \Delta\cP_+^{\rm r}
    &=\frac{|E_0|^2}{2\mu_0ck}\int_{-\frac{\pi}{2}}^{\frac{\pi}{2}} d\theta|f^{\rm r}(\theta)|^2.
    \nn
    \end{align}
For an optical wire described by an optical potential of the form (\ref{v=g}), $f^r(\theta)\approx 0$ for all $\theta\in(-\frac{\pi}{2},\frac{3\pi}{2})$, and $f^l(0)\approx 0$ by virtue of the Reciprocity Principle. Therefore $\Delta\cP_\pm^{\rm r}\approx 0$, while $\Delta\cP_\pm^{\rm l}>0$. This in particular shows that the wire displays nonreciprocal transmission, because $\Delta\cP_+^{\rm l}\neq
\Delta\cP_-^{\rm r}$.

It is interesting to observe that $\Delta\cP_-^{\rm l}$ and $\Delta\cP_+^{\rm l}$ are respectively proportional to the total reflection and transmission cross section, i.e., $\int_{\frac{\pi}{2}}^{\frac{3\pi}{2}} d\theta|f^{\rm l}(\theta)|^2$ and $\int_{-\frac{\pi}{2}}^{\frac{\pi}{2}} d\theta|f^{\rm l}(\theta)|^2$. This in turn implies that the total reflected and transmitted power for $\psi^{\rm l}$ are quadratic functions of the strength of the potential.\footnote{This means that if we scale the potential as $v\to\alpha v$ for a sufficiently small real number $\alpha$ so that the first Born approximation remains valid, the total reflected and transmitted power for $\psi^{\rm l}$ scale by a factor of $\alpha^2$.}

Next, we examine a more realistic situation where we intend to determine the power transmitted to a finite screen placed at a large but finite distance from the wire. To this end we introduce the dimensionless parameters:
    \begin{align}
    &\Delta\hat u:=\frac{\br u\kt-\br {u_\emp}\kt}{\br {u_\emp}\kt},
    &&\Delta\hat \bS:=\frac{\br\bS\kt-\br{\bS_\emp}\kt}{|\br{\bS_\emp}\kt|}.
    \label{Delta-u-S}
    \end{align}
In light of the right-invisibility of the wire, these vanish for a right-incident wave. For a scattering solution corresponding to a left-incident wave, they have the following asymptotic expression (in the limit $r\to\infty$):
    \begin{align}
	&\Delta\hat u\approx (1+\cos\theta)\xi(r,\theta),
	&&\Delta\hat \bS\approx
    \xi(r,\theta)(\bbe_x+\frac{\bbr}{r}),
	\label{u-S=}
	\end{align}
where
	\[\xi(r,\theta):=\RE[e^{-ikx}\psi^l]-1	
    =\RE\left[\sqrt{\frac{i}{kr}}e^{ikr(1-\cos\theta)}f^l(\theta)\right].\]

We can use (\ref{u-S=}) to compute the effect of the wire on the power transmitted to a screen determined by $x=d$, $|y|\leq s/2$ and $|z|\leq s/2$, where $d$ and $s$ are real parameters, and $d$ is much larger than the side lengths of the wire. See Fig.~\ref{fig1}. Because the screen is parallel to the $y$-$z$ plane, the difference between the time-averaged power transmitted to the screen in the presence and absence of the wire is proportional to
    \bea
    \Delta\hat\cP&:=&\frac{1}{s}\int_{-\frac{s}{2}}^{\frac{s}{2}}\!\!dy \,\Delta\hat \bS\cdot\bbe_x\Big|_{x=d}\approx \frac{1}{s} \int_{-\frac{s}{2}}^{\frac{s}{2}} \!\! dy\,\Delta\hat u\Big|_{x=d},
    \nn
    \eea
where the second relation follows from (\ref{u-S=}).

\begin{figure}
	\begin{center}
	\includegraphics[scale=.65]{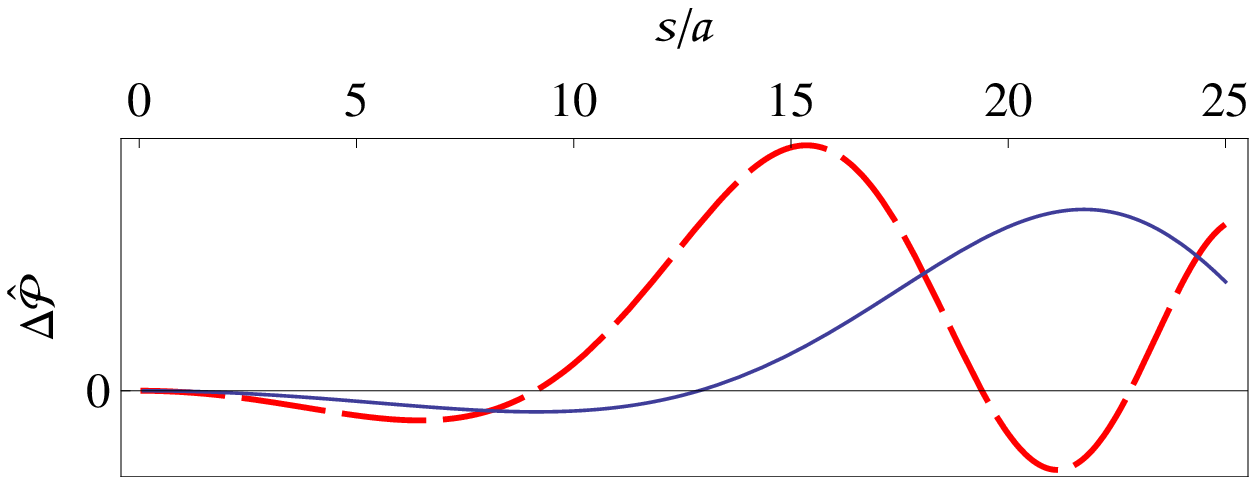} \\[6pt]
    \includegraphics[scale=.65]{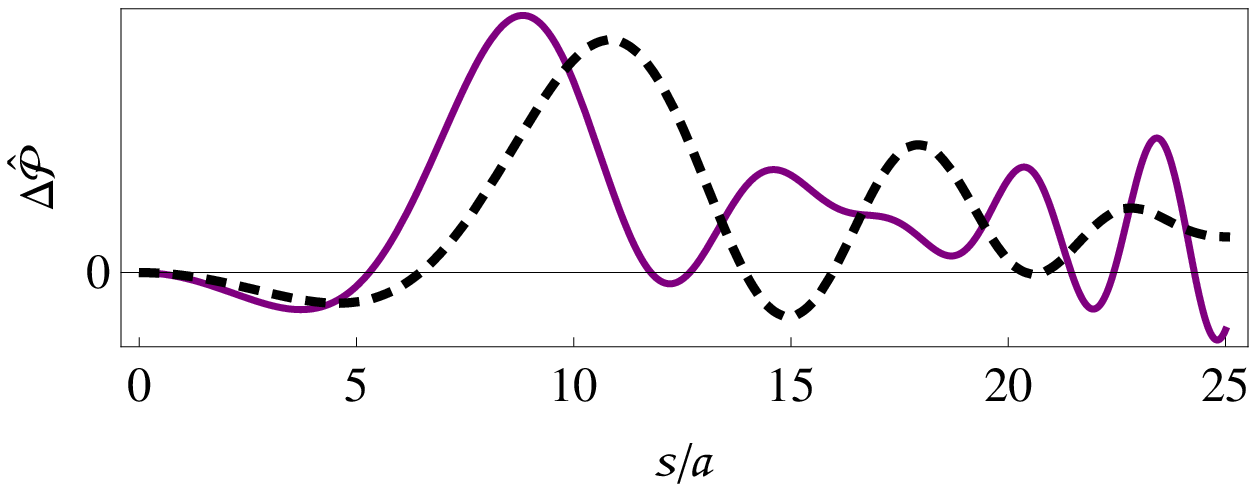}
	\caption{Graphs of $\Delta\hat\cP$ as a function of $s/a$ for $k=2\pi/a$ (thin solid navy curve in the top panel), $k=4\pi/a$ (dashed red curve in the top panel), $k=8\pi/a$ (dotted black curve in the bottom panel), and $k=12\pi/a$ (thick solid purple curve in the bottom panel).}
	\label{fig2}
	\end{center}
	\end{figure}

Figure~\ref{fig2} shows the plots of $\Delta\hat\cP$ as a function of $s/a$ for the cases where $m=-\ell=1$, $g$ is given by (\ref{g=2}) with $b=a$, $d=100 a$, and $k=2\pi/a,4\pi/a,8\pi/a$, and $12\pi/a$. For a left-incident wave the transmitted power exceeds its vacuum value provided that the screen, which is placed to the right of the wire, is sufficiently large. The opposite is the case for a right-incident wave that does not get affected by the wire. This is a clear manifestation of the nonreciprocal transmission property of the system. As seen from Fig,~\ref{fig2}, $\Delta\hat\cP$ tends to zero as $s\to 0$. This is in conformity with the reciprocity principle (Theorem~3),
because $\lim_{s\to 0}\Delta\hat\cP=2\xi(x,0)=2\RE\left[\sqrt{\frac{i}{kx}}f^l(0)\right]$ and $f^l(0)=f^r(\pi)\approx 0$.

The above calculation of the transmitted power to a finite screen reveals its linear dependence on  the strength of the optical potential. This is in contrast to the total transmitted power which has a quadratic dependence on the strength of the potential. In view of the fact that we consider a weak optical potential, so that the first Born approximation is reliable, this shows that the setup involving a finite screen is more desirable for displaying the nonreciprocal transmission property of the system.

\section{Generalization to 3D}
\label{Sec6}

The results we have reported in the preceding sections admit straightforward generalizations to three dimensions. Here we provide a brief summary of the results in this direction. We begin by listing some convenient conventions and notation.

In what follows $v$ is a real or complex scattering potential defined on $\R^3$. We use a Cartesian coordinate system in which the scattering axis corresponds to the $z$-axis. Following \cite{p129} we employ over-arrows to denote two-dimensional vectors obtained by projecting three-dimensional vectors along the $x$-$y$ axis. For example, $\vec\rho:=(x,y)$ and $\vp:=(p_x,p_y)$, whereas $\bbr:=(x,y,z)$ and $\mathbf{p}:=(p_x,p_y,p_z)$. Similarly we employ the notation: $\vd_p:=(\partial_{p_x},\partial_{p_y})$ and $\vec\nabla^2:=\partial_x^2+\partial_y^2$.

\subsection{Transfer Matrix in 3D}
In order to define an appropriate notion of a transfer matrix in three-dimensions, we pursue the approach of Sec.~\ref{Sec2} with the role of $x$, $p$, and $[-k,k]$ respectively played by $z$, $\vec p$, and $\sD_k:=\big\{\vec p\in\R^2\:\big|\:|\vec p|\leq k\big\}$.
In particular the scattering solutions of the Schr\"odinger equation tend to
    \bea
    \frac{1}{(2\pi)^2}
	\int_{\sD_k} d^2p\, e^{i\vec p\cdot\vec\rho}\left[A_\pm(\vec p)e^{i\omega(\vec p)z}+
	B_\pm(\vec p) e^{-i\omega(\vec p)z}\right],~~~
    \label{3D-asymp}
	\eea
as $z\to\pm\infty$, and the transfer matrix $\bM(\vp)$ is defined by (\ref{M-def}) with $p$ changed to $\vp$. Again we can show that it satisfies (\ref{M=}) for an effective Hamiltonian operator of the form 
    \[\bH(z,\vec p):=\frac{1}{2\omega(\vec p)}\: e^{-i\omega(\vec p)z\boldsymbol{\sigma}_3}
	v(i\vec\partial_p,z)\,\boldsymbol{\cK}\,e^{i\omega(\vec p)z\boldsymbol{\sigma}_3},\]
provided that we replace $x$ with $z$, \cite{p129}.


Next, we recall that in three dimensions the scattering solutions of the Schr\"odinger equation admit the asymptotic expression:
    \[e^{\pm ik z}+\frac{e^{ikr}}{r}\:f^{\rm l/r}(\vartheta,\varphi)~~~{\rm for}~~~r\to\infty,\]
where $+$ and $-$ correspond to the left- and right-incident waves, $(r,\vartheta,\varphi)$ are the spherical coordinates of $\bbr$ with $\vartheta$ denoting the polar angle, and $f^{\rm l/r}(\vartheta,\varphi)$ is the scattering amplitude. This turns out to satisfy \cite{p129}:
    \be
    f^{\rm l/r}(\vartheta,\varphi)=
    -\frac{ik\left|\cos\vartheta\right|}{2\pi}T^{\rm l/r}_\pm(k\sin\vartheta\cos\varphi,k\sin\vartheta\sin\varphi),
    \nn
    \ee
where $\pm:={\rm sgn}(\cos\vartheta)$ and $T^{\rm l/r}_\pm$ are given by (\ref{Tm-L}), (\ref{Tp-L}), (\ref{Tm-R}), and (\ref{Tp-R}) with $\delta(p)$ replaced by $2\pi\delta(p_x)\delta(p_y)$. Because the entries of the transfer matrix determine $T^{\rm l/r}_\pm$, it contains all the information about the scattering features of the potential.


\subsection{Reciprocity Principle in 3D}

The argument we use in Sec.~\ref{Sec3} to establish the reciprocity principle in two dimensions can be easily generalized to three dimensions. Here we pursue an alternative approach that also has a two-dimensional analog.

Let  $\psi_1$ and $\psi_2$ be any pair of scattering solutions of the Schr\"odinger equation~(\ref{sch-eq}) in three dimensions, and
    {\small
    \be
    j(z):=\int_{\sD_k}\!\! \frac{d^2p}{4\pi^2}\left[\tilde\psi_1(-\vp,z)
    \partial_z\tilde\psi_2(\vp,z)-\tilde\psi_2(\vp,z)\partial_z\tilde\psi_1(-\vp,z)\right],
    \nn
    \ee}%
where $\tilde\psi(\vp,z):=\int_{-\infty}^\infty dx \int_{-\infty}^\infty dy\,
e^{-i(xp_x+yp_y)}\psi(x,y,z)$ is the Fourier transform of $\tilde\psi(x,y,z)$ over $x$ and $y$. Clearly, $\tilde\psi_j$ satisfy the Fourier transformed Schr\"odinger equation,
    \be
    -\partial_z^2\tilde\psi(\vp,z)+v(i\vd_p,z)\tilde\psi(\vp,z)=\omega(\vp)^2\tilde\psi(\vp,z),
    \label{FT-sch-eq}
    \ee
with $\omega(\vp):=\sqrt{k^2-\vp^{\,2}}$. With the help of (\ref{FT-sch-eq}) we can easily show that $\partial_zj(z)=0$. Therefore $j(z)$ is actually $z$-independent. In particular,
    \be
    j(-\infty)=j(+\infty).
    \label{j=j-zz}
    \ee

Next, we use the asymptotic form (\ref{3D-asymp}) of the scattering solutions to compute $j(z)$ for $z\to\pm\infty$. This gives
    \be
    j(\pm\infty)=\int_{\sD_k}\frac{d^2p}{2\pi^2}\left[-i\omega(\vp)\Delta_\pm(\vp)\right],
    \label{j=3D}
    \ee
where $\Delta_\pm(\vp)$ is defined by (\ref{D=}) with $p$ changes to $\vp$. Imposing (\ref{j=j-zz}) and making use of (\ref{j=3D}), we are led to the three-dimensional analogs of (\ref{rec1}) and (\ref{rec2}) which are equivalent to
$f^{\rm l}(0,\varphi)=f^{\rm r}(\pi,\varphi)$.
This argument establishes the validity of Theorem~3 (Reciprocity Principle) in three dimensions.

\subsection{Unidirectional Invisibility in 3D}

The content of Definitions~1 and~2 apply for scattering potentials in three dimensions provided that we respectively use $\vec p$ and $\sD_k$ in place of $p$ and $[-k,k]$. In particular, reciprocal transmission means that $T^l_+(\vp)=T^r_-(\vp)$ for all $\vp\in\sD_k$ or equivalently $f^l(\vartheta,\varphi)=f^r(\pi-\vartheta,\varphi)$ for all
$\vartheta\in[0,\frac{\pi}{2})$ and $\varphi\in[0,2\pi)$. This is much stronger a condition than the one imposed by the reciprocity principle. In particular, it is possible to construct left-invisible (right-invisible) scattering potentials with nontrivial transmission from the right (left).

We can follow the approach of Sec.~\ref{Sec4} to construct weak unidirectionally invisible potentials that admit a reliable description using the first Born approximation. Similarly to the case of two dimensions, we find that in three dimensions the first Born approximation yields the following analogs of Eq.~(\ref{Tpm-L=}) and (\ref{Tpm-R=}).
    \begin{align}
	&T^{\rm l}_\pm(\vp)\approx\frac{-i}{2\omega(p)}\,\tilde{\tilde v}(\vp,-p_\mp),
	\label{Tpm-L=3}\\
	&T^{\rm r}_\pm(\vp)\approx \frac{-i}{2\omega(p)}\,\tilde{\tilde v}(\vp,p_\pm),
	\label{Tpm-R=3}
	\end{align}
where $\tilde{\tilde v}(\vec\fK,\fK_z):=\tilde{\tilde v}(\fK_x,\fK_y,\fK_z)=
    \int_{\R^3} d^3\bbr e^{-i\boldsymbol{\fK}\cdot\bbr} v(\bbr)$
is the three-dimensional Fourier transform of $v(\bbr)$, and $p_\pm:=k\pm\omega(\vp)$. In particular, for left-invisible and right-invisible potentials, we respectively have
    \bea
    \tilde{\tilde v}(\vp,-p_\pm)&\approx& 0,\\
    \tilde{\tilde v}(\vp,p_\pm)&\approx& 0.
    \label{r-inv-3D}
    \eea

Repeating the construction of Sec.~\ref{Sec4}, we can use (\ref{r-inv-3D}) to construct right-invisible potentials of the form
    {\small
    \bea
	v(x,y,z)&=&\chi_{c}(z)\Big\{
    -\Big(1+\frac{\ell e^{i\ell K x}-m e^{im K x}}{\ell-m}\Big)\vec\nabla^2g(x,y)
	\nn\\
	&& +\frac{\ell m K}{\ell-m}
	\Big[(\ell K-2k)e^{i\ell K x}-(m K-2k)e^{i m K x}\Big]g(x,y)\Big\},
	\nn
	\eea}%
where $c$ is a length scale, $\ell$ and $m$ are distinct nonzero integers, $K:=2\pi/c$, and $g$ is an arbitrary well-behaved function with sufficiently rapid asymptotic decay rate. For example, let $a$ and $b$ be positive real parameters, $g_0$ be a nonzero real or complex number, and
    \[g(x,y)=g_0a^{-4}b^{-4}x^2y^2(x-a)^2(y-b)^2\chi_a(x)\chi_b(y).\]
Then for each choice of $\ell$ and $m$, $v(x,y,z)$ is a right-invisible potential that vanishes outside the (rectangular) cube defined by $0\leq x\leq a$, $0\leq y\leq b$, and $0\leq z\leq c$. It is easy to check that it is neither left-reflectionless nor left-transparent. Therefore, it is a finite-range unidirectionally invisible potential with nonreciprocal transmission.

\section{Concluding Remarks}
\label{Sec7}

For a scattering potential in one dimension, the reciprocity in transmission is a consequence of the $x$-independence of the Wronskian of the left- and right-incident scattering solutions of the Schr\"odinger equation. In this article, we have introduced a similar conserved quantity involving solutions of the Schr\"odinger equation in two and three dimensions and used it to prove a reciprocity theorem that applies for arbitrary real and complex scattering potentials in these dimensions. This theorem states that the forward scattering amplitude for left- and right-incident waves coincide. The condition for having reciprocal transmission in higher dimensions is much stronger. It requires the invariance of the scattering amplitude under the parity transformation that flips the orientation of the scattering axis. In two dimensions it takes the form $f^l(\theta)=f^r(\pi-\theta)$, while the reciprocity principle only demands $f^l(0)=f^r(\pi)$.

In particular, it is possible to have potentials whose total transmission cross section to the left differs from that to the right. An extreme example is a unidirectionally invisible potential that is not transparent from both the directions. We have given a precise definition for the concept of unidirectional invisibility in two and three dimensions and devised a general method of constructing explicit examples of perturbative unidirectionally invisible potentials. These are the multidimensional generalizations of the complex exponential potentials (\ref{exp}) in one dimension whose study initiated the overwhelming recent interest in unidirectional invisibility.

We have constructed an infinite family of unidirectionally invisible potentials in two dimensions that admit simple optical realizations. These include as a special case a right-invisible potential modeling an optical wire with a rectangular cross section that is neither reflectionless nor transparent from the left. For this model we have explored the behavior of the total transmitted and reflected power as well as the power transmitted to a finite screen placed at finite distance from the wire. We have also constructed similar unidirectionally invisible potentials in three dimensions.

We expect our results to pave the way for a systematic study of the phenomenon of unidirectional invisibility in realistic multidimensional systems. They should be of particular interest for devising optical and acoustic devices displaying nonreciprocal transmission. \vspace{6pt}

\subsection*{Acknowledgments} We are indebted to Aref Mostafazadeh for his help in preparing Fig.~\ref{fig1}. This work has been supported by the Scientific and Technological Research Council of Turkey (T\"UB\.{I}TAK) in the framework of the project no: 112T951 and by Turkish Academy of Sciences (T\"UBA).

\ed

\section*{Appendix: Calculation of $\Delta\hat u $ and $\Delta\hat \bS$}

The time-averaged energy density and Poynting vector of the electromagnetic waves propagating in a charge-free, nonmagnetic, isotropic medium are respectively give by
    {\small\begin{align}
    &\br u\kt=\frac{1}{4}\RE(\bE\cdot\bD^*+\bB\cdot\bH^*),
    &&\br\bS\kt:=\frac{1}{2}\RE(\bE\times\bH^*),
    \label{u-S}
    \end{align}}%
where $\bE$ and $\bH$ are the electric and magnetic fields, $\bD=\varepsilon_0\hat{\varepsilon}\,\bE$, $\bB=\mu_0\bH$, and $\varepsilon_0$ and $\mu_0$ are the permittivity and permeability of the vacuum, respectively. For time-harmonic waves,
we can use Maxwell's equations to establish $\bH=\frac{1}{ik}\sqrt{\frac{\varepsilon_0}{\mu_0}}\boldsymbol{\nabla}\times\bE$. Substituting this relation in (\ref{u-S}) and making use of (\ref{E=ez}) and (\ref{psi-left}), we can express $\br u\kt$ and $\br\bS\kt$ in terms of the scattering amplitude $f^{\rm l}(\theta)$. In view of (\ref{Delta-u-S}) this gives (\ref{u-S=}).